\documentclass{aa}  
\usepackage{graphicx}
\usepackage{graphicx}
\usepackage{natbib}
\bibpunct{(}{)}{;}{a}{}{,}
\newcommand{\mum}{$\mu$m~}

\newcommand{\Vt}{$v_{\rm t}$}

\newcommand{\Tef}{\mbox{$T_{\rm eff}$}}

\newcommand{\vunit}{\mbox{\,km\,s$^{-1}$}}
\newcommand{\mic}{\mbox{$\,\mu$m}}

\newcommand{\ltsimeq}{\raisebox{-0.6ex}{$\,\stackrel
        {\raisebox{-.2ex}{$\textstyle <$}}{\sim}\,$}}

\newcommand{\vmon}{\mbox{V838~Mon}}

\begin{document}

\title{The properties of \vmon\ in 2002 November}

\author{
Ya. V. Pavlenko\inst{1,2}
\and J. Th. van Loon\inst{3}
\and A. Evans\inst{3}
\and M. T. Rushton\inst{3}
\and B. M. Kaminsky\inst{2}
\and A. V. Filippenko\inst{4}
\and R. J. Foley\inst{4}
\and W. Li\inst{4}
\and B. Smalley\inst{3}
\and L. A. Yakovina\inst{2}
}
\institute{Centre for Astrophysics Research, University of Hertfordshire, 
College Lane, Hatfield, Hertfordshire AL10 9AB, UK 
\and Main Astronomical Observatory, Academy of Sciences of the Ukraine, Golosiiv
     Woods, Kyiv-127, 03680 Ukraine
\and Astrophysics Group, School of Physical 
   \& Geographical Sciences, Keele University, Keele
     Staffordshire, ST5 5BG, UK
\and Department of Astronomy, 601 Campbell Hall, University of California, Berkeley,
     CA 94720-3411, USA
}

\offprints{Ya. V. Pavlenko}
\mail{yp@mao.kiev.ua}

\date{Version of 15 Mar 2006}

\authorrunning{Pavlenko et al.}
\titlerunning{\vmon}

\abstract{We present the results of modelling the 0.45--1\mic\ spectral energy
distribution of \vmon\ for 2002 November. Synthetic spectra were calculated
using the NextGen model atmospheres of Hauschildt et al. (1999), which incorporate
line lists for H$_2$O, TiO, CrH, FeH, CO, and MgH, as well as the VALD atomic line list.
Fits to the observed spectra show that, in 2002 November, the effective temperature
of \vmon\ was approximately $2000\pm100$~K. Our theoretical spectra show a
comparatively weak dependence on $\log g$. Preliminary analysis of the hot star 
observed together with \vmon\ shows it to be a normal B3V dwarf.
}

\maketitle

\section{Introduction}

\vmon\ (also known as  EQJ0704.0-0350, GSC~04822-00039, IRAS~07015-0346,
USNO-A2.0 0825-03833116, AAVSO 0659-03) has been the subject of
intense interest since its discovery as an eruptive variable by
\cite{brown} on January 6, 2002. Two further  maxima on the light curve
  subsequently
occurred, in 2002 February (\citealt{Munari2002, Kimeswenger2002,
crause2003}) and the optical flux in the $V$ band increased by
$\sim 9$~mag. The luminosity peaked at $V \approx 7$ in 2002 February.
Later, a gradual decline in the $V$ flux began which, by 2003 January,
had faded by 8 mag.

\cite{Henden2002}
% Henden et al. (2002)
reported the presence of a light echo, subsequently studied in detail
by \cite{Bond2003} and \cite{crause2005}. The reflecting dust may be
interstellar \citep{Tylenda2005a} or may have originated in the
envelope of \vmon\ and ejected in the past \citep{loon2004}.
Van Loon et al. (2004) reported the discovery of the multiple dust shells
around \vmon.  Alternatively, these shells arose through heavy mass loss
during the evolution of a massive progenitor (Munari et al. 2005). 
 
The distance to \vmon\ can be estimated from the evolution of the
light echo \citep{Henden2002}. According to works based on
Hubble Space Telescope data, its distance is $> 6$~kpc \citep{Bond2003}, 
 $8\pm 2$~kpc
\citep{Tylenda2004}, or as high as 12 kpc (Crause et al. 
2005). However, according to recent work based on HST data, its distance is 
5.9 kpc (Sparks et al. 2006). 
If these estimates are correct, then at the time
of maximum brightness \vmon\ was the most luminous star in the Galaxy.

The cause of the eruption of \vmon\ and the nature of its progenitor
are unclear. The integrated colors of the progenitor looked that
of an unreddened F-type star \citep{Munari2002}. 
%%\cite{desidera2002} reported a
%%faint hot continuum at short wavelengths, recently confirmed by
%%\cite{munari05} and identified as a B3V companion.
\cite{desidera2002} discovered spectroscopically a hot companion
    to the outbursting star, later confirmed by Wagner et al. (2002), and
    classified as a B3V star by \cite{munari05}.

The spectral, photometric, and polarimetric evolution of \vmon\ has been
described in several studies (\citealt{Munari2002, Kimeswenger2002,
Kolev2002, Wisniewski2003, Osiwala2003, crause2003, Kipper2004, rushton05}).
By mid-April 2002, bands of TiO had appeared in the spectrum and by May,
the spectrum had evolved to a ``very cold'' M giant \citep{banerjee02}. Its
infrared spectrum in 2002 October was characterized as an ``L-supergiant''
by \cite{evans03}.

The complexity and the rarity of V838 Mon optical spectrum resulted in
    only a few sporadic attempts to derive the star's atmospheric properties.
%%Very little has been done on a fine analysis of \vmon's spectrum.
\cite{Kipper2004} found that, for the iron group of elements,
$\mbox{[M/H]} = -0.4$, while abundances of lithium and of some
s-process elements are clearly enhanced. Later, \cite{kaminsky05}
provided fits to echelle data and obtained a similar result,
$\mbox{[Fe/H]} = -0.3\pm0.2$. These results were obtained using a
static LTE model and are therefore very dependent on the model
atmosphere and spectrum synthesis assumptions.

In this paper we discuss fits to the spectral energy distribution of
\vmon\ using data obtained in 2002 November. The observational data
used in this paper are described in Section 2. Section 3 explains
some of the background to our work and some details of the procedure
and input parameters used. The results are discussed in Section 5.

\section{Observations}

The spectrum we consider was obtained on 2002 November 6 with the Kast
spectograph on the Cassegrain focus of the Shane 3-m telescope at Lick
Observatory. The resolution of the spectrum is 
$\lambda/\Delta\lambda=3,000$.
The observing and data reduction processes 
are the same as those in \cite{rushton05} and are not repeated here.

\section{Computation of theoretical spectra}

Theoretical spectral energy distributions (SEDs) were computed for
model atmospheres of giants from the NextGen grid of
\cite{hauschildt99}, with solar metallicity \citep{anders89}.

In our computations we used a number of model atmospheres, with
\Tef = 2000--2200~K and $\log g = 0, 0.5$. Computations of the
synthetic spectra were carried out using the program WITA6
\citep{pavlenko00}, assuming LTE and hydrostatic equilibrium for a
one-dimensional model atmosphere without sources and sinks of
energy.

The equations of ionization-dissociation equilibrium were solved
for media consisting of atoms, ions, and molecules. We took into
account $\sim100$ components \citep{pavlenko00}. The constants for
the equations of chemical balance were taken from \cite{tsuji73}.
Molecular line data were taken from a variety of sources:
\begin{enumerate}
\item [{(i)}] the TiO line lists of \cite{plez98};
\item [{(ii)}] CN lines from CDROM 18 \citep{kurucz93};
\item [{(iii)}] CrH and FeH lines from \cite{burrows02} and
      \cite{dulick03}, respectively;
\item [{(iv)}]  lines of H$_2^{16}$O \citep{bt2};
\item [{(v)}] absorption by VO, and by a few molecules of (in the case of
      \vmon) lesser importance, was computed in the JOLA
      approximation (see \citealt{pavlenko00}); and
\item [{(vi)}] atomic line list from VALD \citep{kupka99}.
\end{enumerate}

\begin{figure*}
\includegraphics [width=160mm]{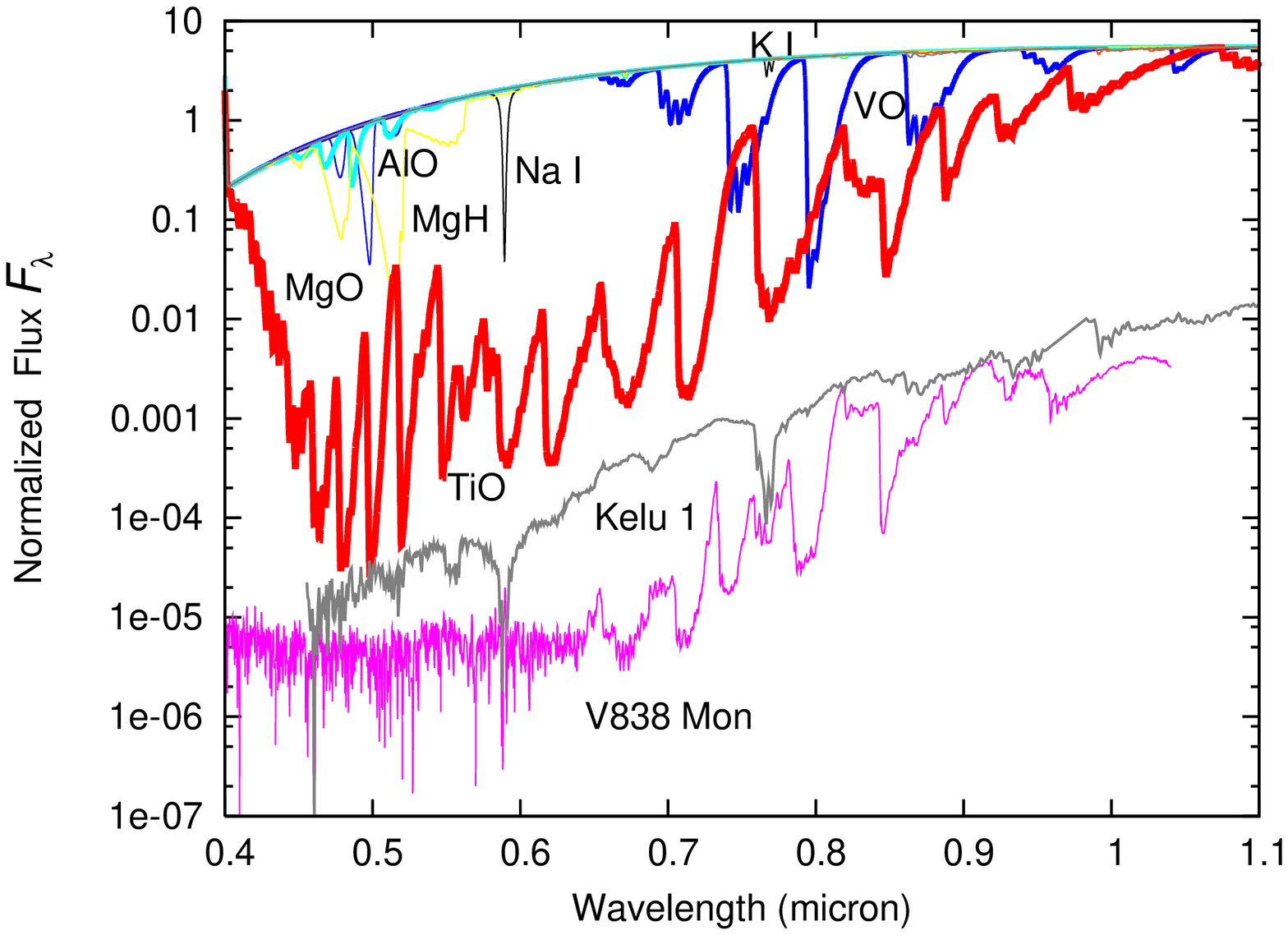}
\includegraphics [width=160mm]{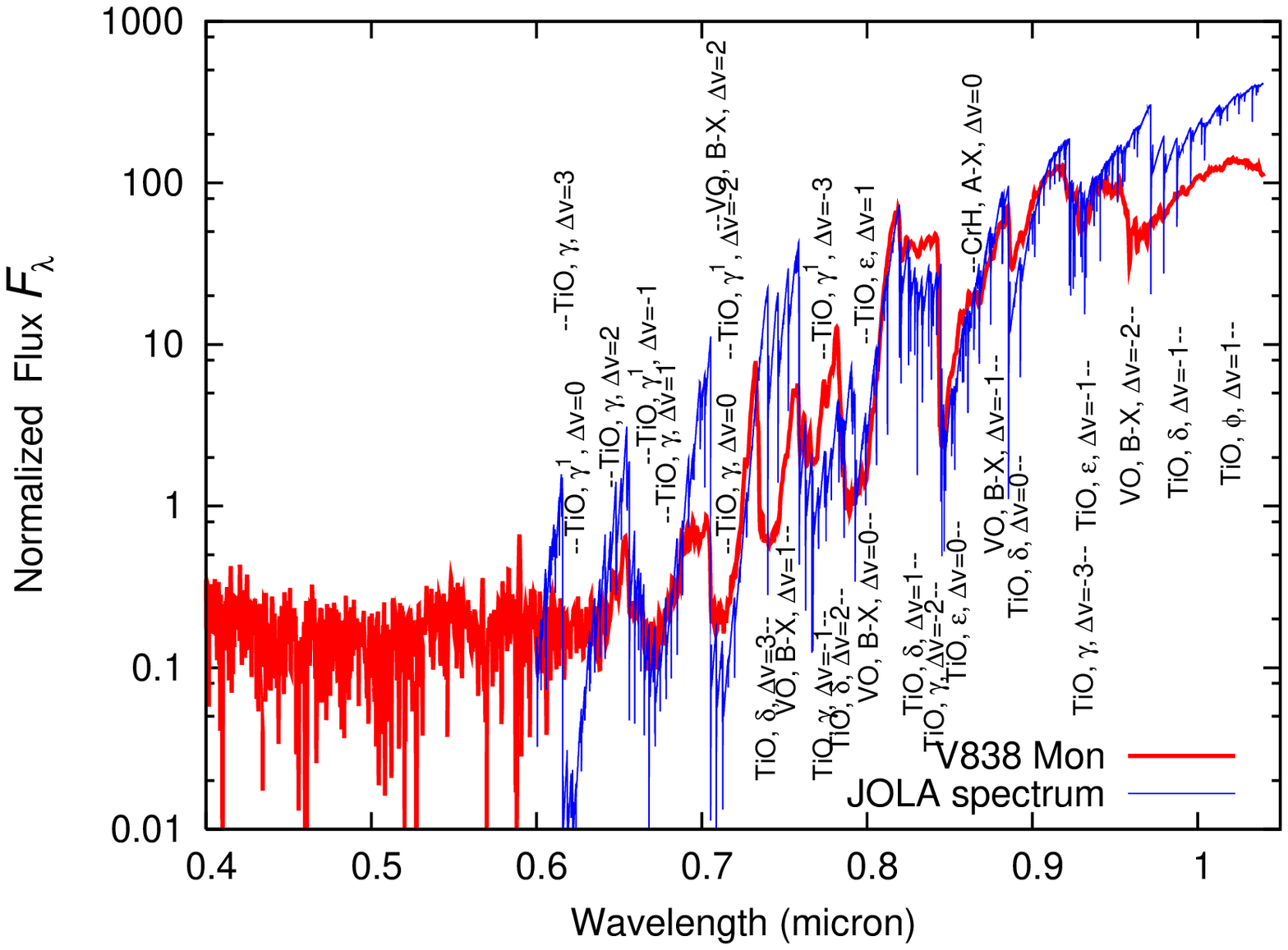}
\caption{Top: contribution of different molecules to the
formation of the spectrum of \vmon. The two lower spectra represent
observed fluxes of L-dwarf Kelu1 \citep{leggett02} and \vmon.
Bottom: Identification of the main molecular features in the
spectrum of \vmon. {\em Data for the figure plotting 
and color version of the plots are available on
ftp://ftp.mao.kiev.ua/\%2f/pub/users/yp/Fig.1}} \label{opacities}
\end{figure*}

The profiles of molecular and atomic lines are determined using
the Voigt function $H(a,v)$. Their natural ($C_2$) and van der Waals
broadening ($C_6$) parameters are taken from \cite{kupka99} or, in
their absence, computed following \cite{unsold55}. Owing to the
low temperatures in M-star atmospheres, and the consequent low
electron densities, Stark broadening may be neglected; on the
whole the effects of pressure broadening dominate. Computations
for synthetic spectra were carried out with a step of 0.003\mic,
for microturbulent velocity $v_{\rm t} = 5\vunit$.

There are several reasons to suspect that dust might be present in the
atmosphere of \vmon, 
see discussion in \cite{banerjee02,evans03,rushton05a}.
However,
examination of the results of preliminary runs of our models without
dust show that the density of condensibles (such as X, Y, Z) at the
appropriate distance is low, so that precipitation in the 
photosphere is extremely
unlikely. Hence, we do not include dust in our models.

In addition, we do not take account of the effects on the model
atmosphere of the presence of a B3 star \citep{munari05}. Over
the wavelength range we are modelling (6000-10\,000~\AA), the SED
is dominated by the cool star and the contribution of the B3 star
is negligible. There may be irradiation effects to be taken
into account if the hot component is part of a \vmon\ binary system,
but we do not pursue this here. We will, however, include the effects of
the B3 star to the overall SED after fitting the red end of the
spectrum (see Section~\ref{hot} below).

The relative importance of the different opacities contributing to
our synthetic spectra is shown in Fig. \ref{opacities}.

\section{Results}

\subsection{Dependence of fits on input parameters}
\label{input}

\begin{figure*}
\includegraphics [width=88mm]{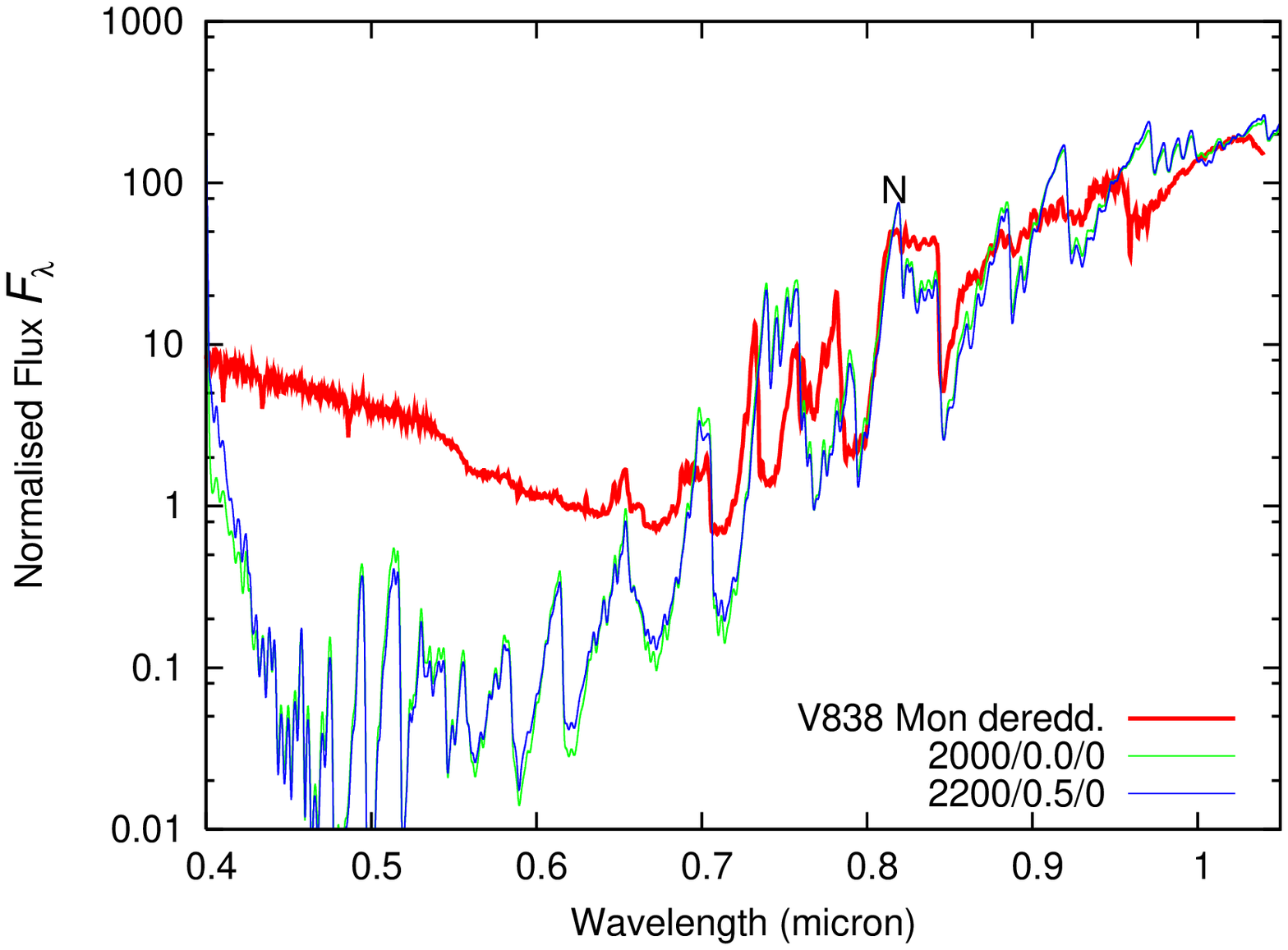}
\includegraphics [width=88mm]{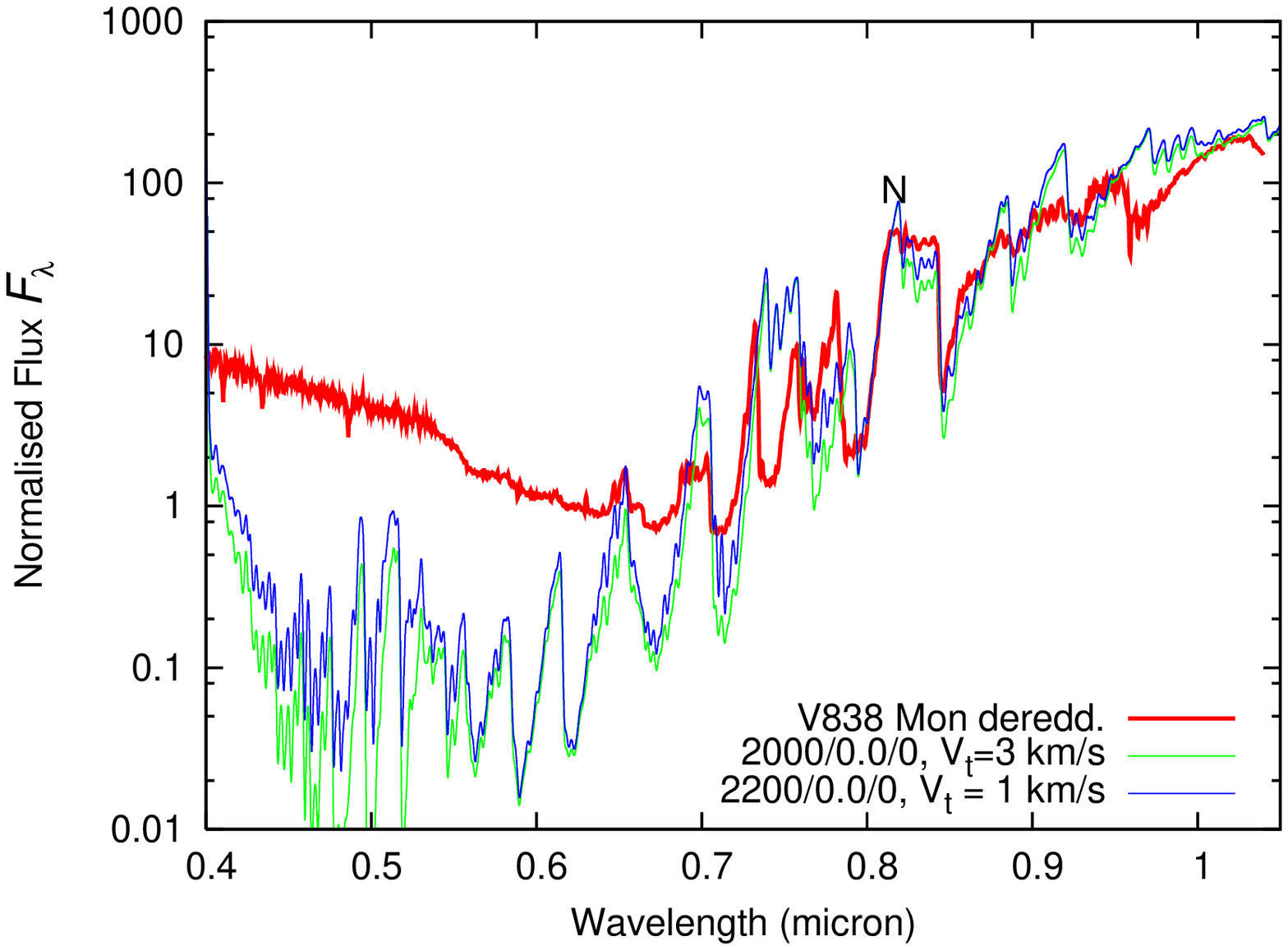}
\includegraphics [width=88mm]{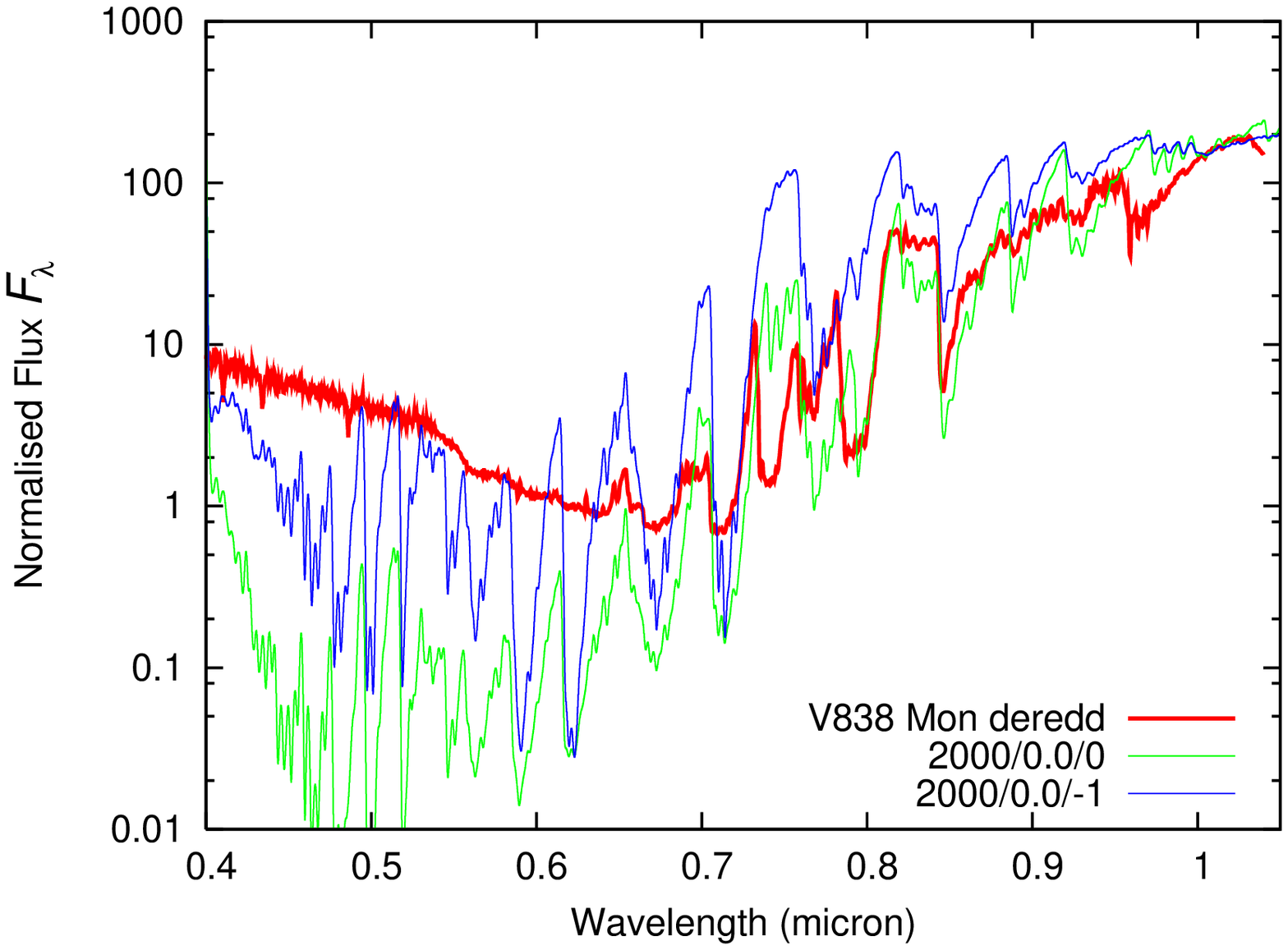}
\includegraphics [width=88mm]{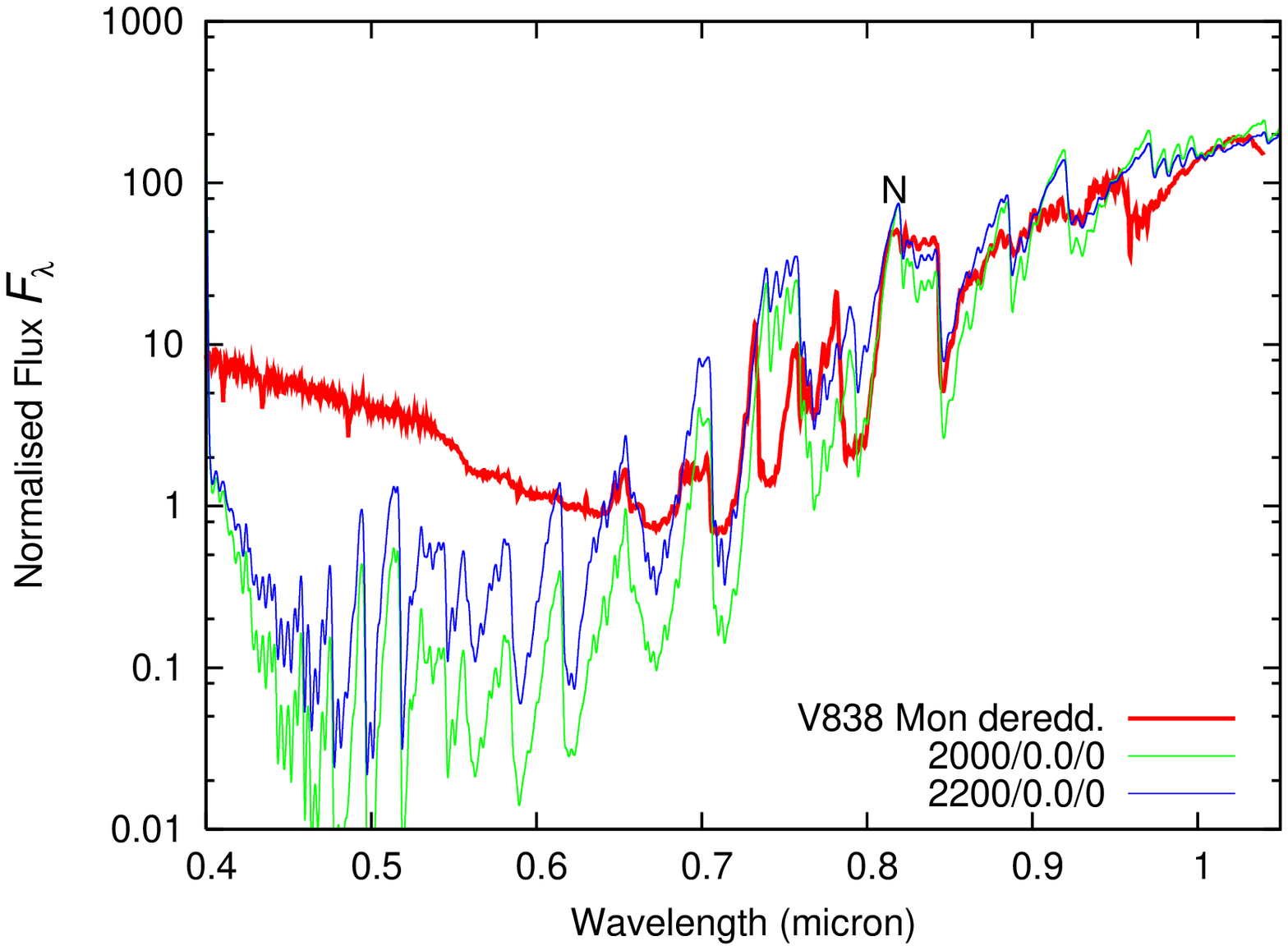}
\caption{Dependence of computed spectra on the gravity (top left),
microturbulent velocity (top right), metallicity (bottom left),
and effective temperature (bottom right). The dereddened observed
spectrum of \vmon\ is shown by the boldfaced line.  Model atmospheres are 
labeled on the figure as \Tef/log/[M/H], i.e 2000/0.0/-1 means
\Tef = 2000, log g =0.0, [M/H] = -1.} \label{_dep}
\end{figure*}

\begin{figure*}
\includegraphics [width=160mm]{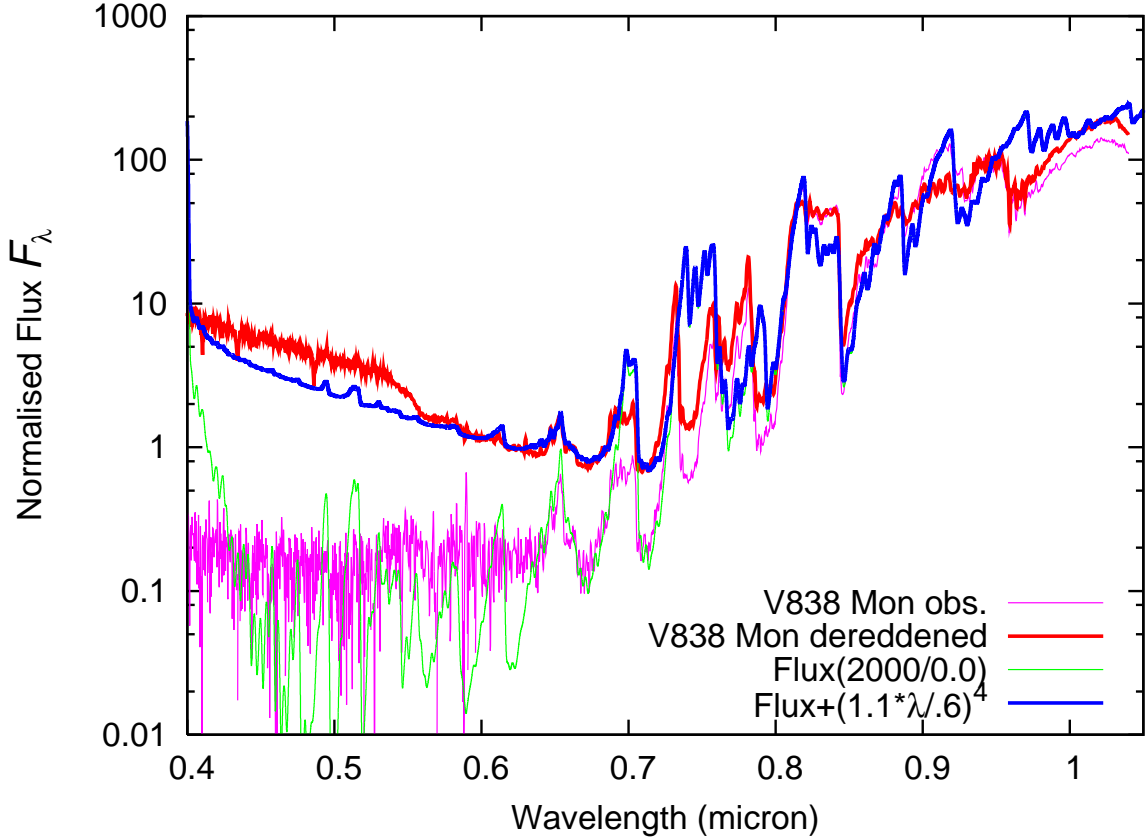}
\caption{Fit of theoretical fluxes to the observed spectrum 
of \vmon. Observed 
and dereddened spectra are shown by solid thin and thick lines, respectively.
The computed \vmon\ spectrum is shown by thin dashed line. 
The combined  
spectrum (computed V838 Mon + Rayleigh-Jeans contribution from the B star) 
is shown by the thick dahsed line.
{\em The color version of the plot is available on
ftp://ftp.mao.kiev.ua/\%2f/pub/users/yp/Fig.3}\label{_fit}}
\end{figure*}
\begin{figure*}
\includegraphics [width=88mm]{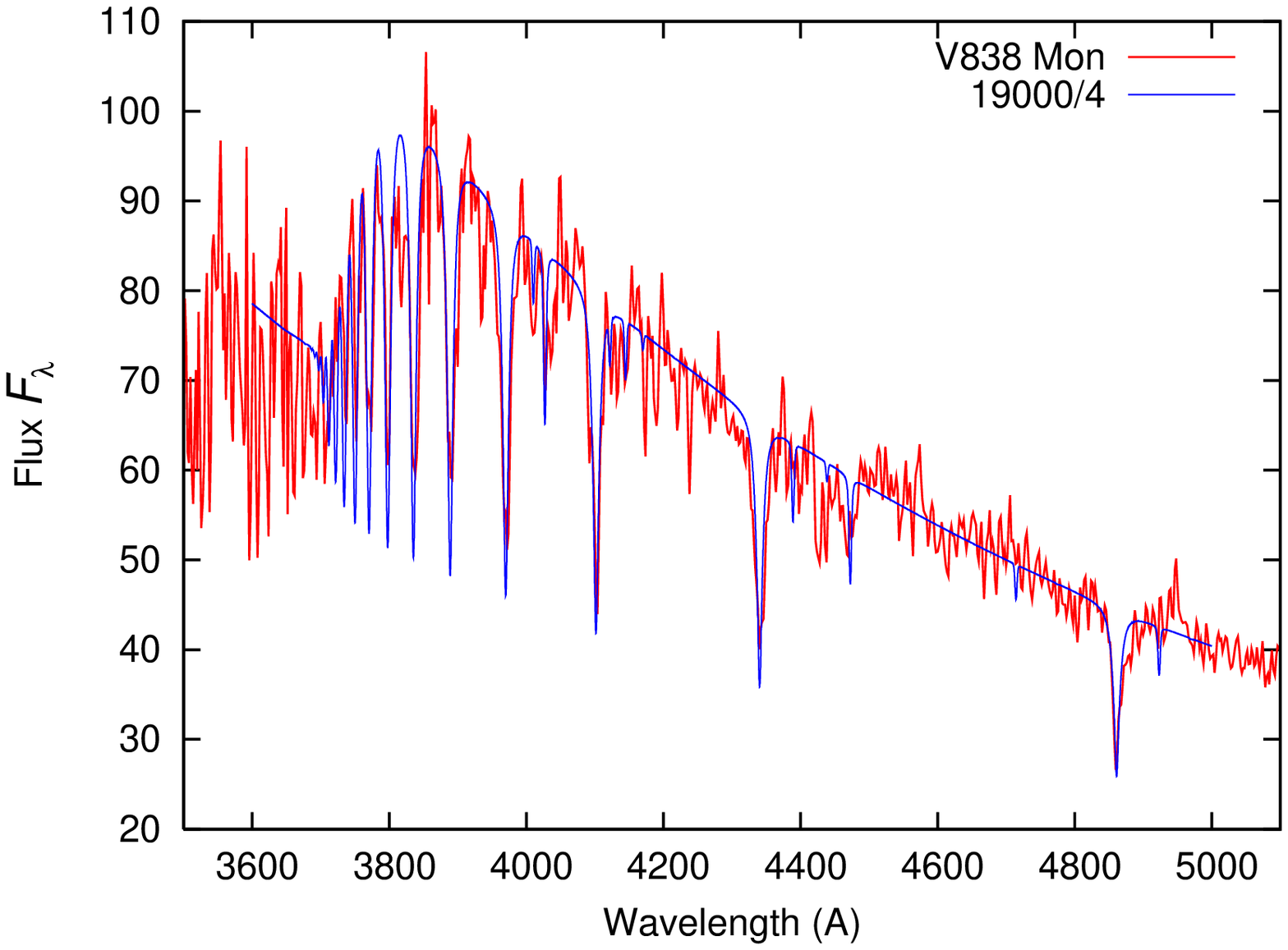}
\includegraphics [width=88mm]{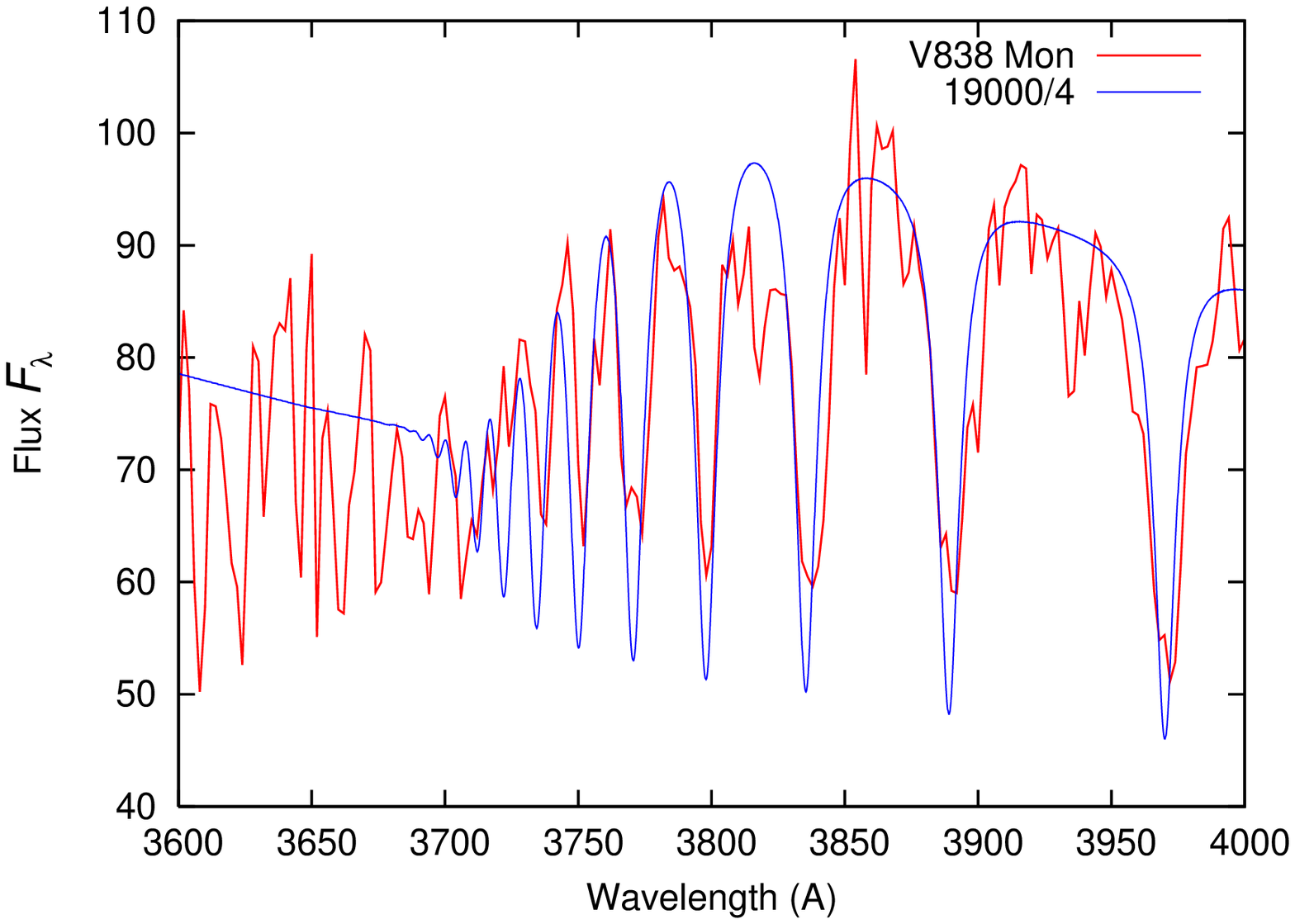}
\caption{Left: Fit of the theoretical spectrum with $\Tef=19000$~K and 
$\log g=4.0$ to the hot companion of \vmon.
Right: The Balmer jump region of the hot companion spectrum. To 
simplify the picture we exlude the metal line absorption 
from the accounted opacity sources.} \label{B3}
\end{figure*}

Changes in our input
parameters, i.e. \Tef, log g, [Fe/H], \Vt, affect the theoretical 
spectral distributions (see Fig. \ref{_dep}) in different
ways, as follows:
\begin{enumerate}
\item [{(i)}] The response of the spectrum to changes in $\log g$ is
relatively weak due to the absence of strong atomic lines with extended
wings.  Due to low densities in the atmosphere of \vmon\ pressure
broadening cannot be significant. Saturated  TiO and VO band heads show
rather low sensitivity on log g. 
\item [{(ii)}] Due to the same reason, changes in the microturbulent 
velocity mainly affect the
heads of strong molecular bands.
\item [{(iii)}] On the other hand, changes in metallicity reduce the
molecular absorption over the entire spectral region.
\item [{(iv)}] Due to the low temperatures in the atmosphere of \vmon\
changes in the effective temperature affect the slope of the 
computed spectra in the region $0.6 - 1$\mic. This effect can be
used to determine the effective temperature of \vmon\ (see
Section~\ref{_res}).
\end{enumerate}

Comparison of the computed spectrum with that of \vmon\ in 2002 November
suggests that the observed fluxes at $\lambda >$0.9 
and $\lambda <$0.8 \mum 
are better fitted with a
slightly metal-deficient atmosphere ($\mbox{[Fe/H}=-0.5$).
This is in  
agreement with the analysis by Kipper et al. (2004) and Kaminsky \&  
Pavlenko (2004).

\subsection{Comparison with the spectrum of Kelu1 (L2)}

We compare (see Fig.~\ref{opacities}) the observed spectrum of \vmon\ with
that of the brown dwarf Kelu1 \citep{leggett02}. The L2 brown dwarf Kelu1 has
approximately the same effective temperature as \vmon\ but, by definition,
its $\log g$ is very much higher. There are several significant points to note:
\begin{enumerate}
\item[{(i)}]  Heads of VO bands are clearly seen on at 0.74, 0.79, 0.87 
and 0.96 \mic.
\item [{(ii)}] The molecular bands of TiO and VO in the spectrum
of \vmon\ are very intense. Due to the low densities these species are practically undepleted
in the atmosphere of \vmon\ despite the low temperatures. On the other hand,
Ti and V containing species are completely depleted in the atmospheres of L-dwarfs.
\item [{(iii)}] The pressure broadened K I and Na I lines are prominent
in the optical spectra of L~dwarfs \cite{pavlenko00}. 
 Instead of L-dwarfs spectra the alkali lines are
absent or very weak in the spectrum of \vmon.
\item [{(iv)}] The electronic bands of CrH and FeH are much weaker in the
spectrum of \vmon\ than they are in Kelu1.
\end{enumerate}

\subsection{The effective temperature of \vmon\ in 2002 \label{_res}}

A comparison of the observed spectrum of \vmon\ with theo\-retical spectra
computed using NextGen model atmospheres is presented in Fig.~\ref{_fit},
in which we use a logarithmic scale for the purpose of illustration as
the observed fluxes cover several orders of magnitude.
The instrumental broadening was modelled by Gaussian profiles set
to the resolution of the observed spectra (FHWM = 2~\AA). The spectrum was
dereddened by $E(B-V)=0.87$ mag \citep{munari05}.

The following are worth noting:
\begin{enumerate}
\item [{(i)}] We can fit the overall slope of the spectrum in the
near-IR/optical region, together with most of the spectral features.
The spectrum is dominated by TiO bands, although some bands of VO are
also prominent.
\item [{(ii)}] This provides clear evidence that, on 2002 November 6,
$\mbox{C/O} < 1$ in \vmon.
\item [{(iii)}] The dependence on $\log g$ is rather weak, as expected
(see Section~\ref{input}). Strong atomic lines are practically absent
in the spectrum.
\item [{(iv)}] The dependence on \Tef\ is much stronger. From comparison of
the computed and observed SEDs, we conclude that the effective temperature
of \vmon\ was $\Tef = 2000\pm100$~K in 2002 November.
\end{enumerate}

Our value of \Tef\ for 2002 November 6 compares favourably with that
determined by \cite{tylenda05} for 2002 day~301 (October 28, the nearest
in Tylenda's compilation); he obtained $\Tef=2180$~K 
(formally this corresponds to a spectral class M8.5).

\subsection{The hot component in the \vmon\ spectrum}
\label{hot}

Our observed spectrum of \vmon\ shows a rise in the flux toward the
blue end of the spectrum, at $\lambda\ltsimeq 6500$~\AA, which can be
attributed to the presence of the ``hot'' star; this may be a background
or foreground object, or even a member of a \vmon\ binary system
(see discussion in \citealt{munari05}).  
It is worth noting that all the Balmer hydrogen lines, from H$\beta$ to
the Balmer jump, are clearly seen in the observed spectrum. 

We first model the contribution of the hot component by a simple
Rayleigh-Jeans law $F_{\lambda} = \mbox{const}/{\lambda}^4$ 
and normalise both fluxes to the
same value at 0.67\mic. This comparatively simple procedure
allows us to fit to the observed spectrum over a wide spectral
range, from the blue to the near-infrared.

We next computed a set of theoretical spectra using the standard
Kurucz model and WITA6, for different values of \Tef\ and $\log g$. 
A comparison of the observed spectrum of the hot companion with
theoretical spectra is presented in Fig.~\ref{B3}. 
Unfortunately, our low-resolution spectrum, with its low signal-to-noise
ratio in the blue, cannot provide a reliable estimate of the atmospheric
parameters (effective temperature, gravity and/or metallicity) of the hot
component. From the fit to the observed spectrum we estimate  
$\Tef=18000 - 23000$~K and $\log g=4.0$.

From evolutionary calculation \citep{schaller92} 
we find the mass of the B star; we then
computed the possible range of spectral parallaxes for our 
hot component.  
We obtain $M = 6~M_\odot$, d = 7.2~kpc for \Tef\ = 18000 and 
$M = 9~M_\odot$, d = 14.4~kpc for \Tef\ = 23000 K, respectively, 
i.e only in the case of 
 \Tef\ $<$ 21000 K the 
hot component can be a member of the \vmon\ binary system.

%%Computed spectral parallaxes show that, in the case of \Tef = 23000~K,
%%the spectral companion is likely a background star. However if \Tef\
%%for the hot star is $\sim18000 - 19000$~K then the hot component might
%%form a physical binary with \vmon. 

\section{Discussion and conclusions}

It seems that the pseudo\-photosphere of \vmon\ was evolving
sufficiently slowly that its optical spectrum could be approximated by a
sequence of classical,  i.e. static model atmospheres 
\citep{kaminsky05}; we find that
the standard computational procedures for SEDs, applied to \vmon, give
results that are reliable enough to determine the effective temperature.
 It is worth noting that the presence of P Cyg 
profiles indicates expansion velocities up to 200\vunit. However the 
emission components of the P Cyg profiles form at the outer boundary of  
the envelope, well above the region where the observed continuum and 
absorption lines form.

There are distinct differences between the SED of \vmon\ and of
late-type dwarfs with similar spectral type, which obviously arise from
their differing physical nature and the input physical data. In the case
of \vmon\ we have modelled its SED at an advanced stage of its evolution
and find that its optical spectrum is formed by absorption of the saturated
bands of VO and TiO.

We note that our models do not satisfactorily reproduce the absorption
band located around 0.75\mic; this absorption cannot be identified as TiO
or VO bands (see Fig.~\ref{opacities}). Hypothetically, the strong band
of CH$_4$ could be formed in a low-temperature regime at these wavelengths.
This band, as well as the bands at 0.619 and 0.890\mic, are known to be
present in the spectra of Jupiter and other gas giants. Possibly the band
may be formed in the outermost layers of the extended envelope of \vmon.
We plan to provide a more detailed 
study of the molecular features in a forthcoming paper.

Nonetheless, we have obtained a qualitatively good fit of the spectrum
of \vmon\ over a wide spectral range (0.4--1\mic) with a model atmosphere
having $\Tef=2000 \pm 100$~K and C/O $<$ 1, and a hot component 
consistent with Teff=18000 - 23000 K and log g =4.0 $\pm$ 0.5; 
this agrees well with previous studies \citep{tylenda05}.
The cool component is slightly better 
reproduced with a moderately sub-solar metallicity model.

%%Moreover the computation of the SED was carried out within the framework
%%of the classical approach: there is not at this stage any need to include
%%additional physical processes to find the best fit to the observed SEDs.

%%Unfortunately, from our study we cannot determine other basic parameters,
%%i.e \Vt, log g and [Fe/H]. The main features in our low resolution spectra
%%are formed by the heads of saturated molecular bands. These spectral features show 
%%rather low sensitivity on small variations of \Vt, log g and [Fe/H].
%%They can be determined from the fits to high resolution spectra.

%%Our study enables us to draw the following conclusions:
%%\begin{enumerate}
%%\item [{(i)}] We can use fits to SEDs of \vmon\  to determine its basic
%%      parameters; in particular the effective temperature $\Tef=2000\pm100$~K.
%%\item [{(ii)}] We clearly see that, in 2002 November 6, $\mbox{C/O} < 1$ in
%%      the atmosphere of \vmon;
%%\item [{(iii)}] The contribution of the hot star to the spectrum of \vmon\
%%      is clearly seen in 2002 November; this hot component is consistent
%%      with $\Tef=19000$~K and $\log g=4.0$.
%%\end{enumerate}

\section{Acknowledgments}

YP's work was partially supported by the Royal Society and the
Leverhulme Trust. The computations were in part performed using
the resources of HiPerSPACE computing facility at UCL which is
part funded by the UK Partical Physics and Astronomy Research
Council (PPARC). AVF acknowledges support from USA NSF Grant
AST-0307894. We thank anonymous referee for the helpful comments.

\end{document}